\documentclass[showpacs,aps,pre,twocolumn,superscriptaddress]{revtex4-2}

\usepackage{amssymb}
\usepackage{amsmath}
\usepackage[sort&compress]{natbib}
\usepackage{graphicx}
\usepackage{color}
\usepackage{CJK}
\usepackage{dcolumn}   
\usepackage{hyperref}

\usepackage{setspace}

\begin{document}
		
		\title{Global self$-$similarity of dense granular flow in hopper: \\ the role of hopper width}
		
		\author{Changhao Li} \author{Xin Li} \author{Xiangui Chen} \author{Zaixin Wang} \author{Min Sun}
		\author{Decai Huang}
		\email[Corresponding author: ] {hdc@njust.edu.cn}
		\address{Department of Applied Physics, Nanjing University of Science and Technology, Nanjing 210094, China}
		
		\date{\today}
		
		\begin{abstract}
			
			The influence of hopper width on dense granular flow in a two$-$dimensional hopper is investigated through experiments and simulations. Though the flow rate remains stable for larger hopper widths, a slight reduction in hopper width results in a significant increase in flow rate for smaller hopper widths. Both Beverloo\('\)s and Janda\('\)s formula accurately capture the relationship between the flow rate and outlet size. Flow characteristics in the regions near the outlet exhibit local self$-$similarity, supporting Beverloo and Janda's principles. Moreover, global self$-$similarity is analysed, indicated by the transition in flow state from mass flow in regions far from the outlet to funnel flow near the outlet. The earlier occurrence of this transition favors to enhance the grain velocity and consequently increases the dense flow rate. An exponential scaling law is proposed to describe the dependencies of flow rate, grain velocity, and transition height between the mass flow pattern and the funnel flow pattern on silo width.
			
		\end{abstract}
		
			
	\maketitle
	\section{Introduction}
	\label{Intro}
	
	Granular flow exists extensively in nature and industrial processes, attracting significant attention from the fields of science and engineering\cite{Jaeger1996,Aranson2006,Iverson1997,Kamrin2012,ChenCES2022,RauterNature2022}. It exhibits complex flow behaviors, such as the segregation of binary mixtures\cite{Savage1988,Gray2021,LueptowCES2023}, the transition between dilute state and dense state\cite{HouPRL2003,Huang2006PRE,Huang2011PLA,HuPRE2015,HuangPT2021}. Gravity$-$driven granular flow in a hopper serves a good example where various intrinsic flow properties can be observed. Numerous theoretical and technological efforts have been focused on studying this special process because of its ubiquity \cite{Hagen1852,Brown1961Nature,Beverloo1961,GoldhirschGM2010,HiltonPRE2011,ZuriguelPRL2015,ZuriguelPRE2017,ArevaloPRE2022,YangPT2022,BhatejaPOF2023,MazaPRE2024}. 
	How to control the flow rate and understanding the governing principles are fundamental and critical issues. 
	
	In classical research, differential relationship between the physical quantities is the key point to define the dynamical characteristics.  Given this consideration, Beverloo law is a successful empirical formula describing the relationship between the dense flow rate $Q$ and the outlet size $D$ by using a simple physical analysis.\cite{Beverloo1961}
	\begin{equation}
		Q=C{\sqrt{g^{*}}}(D-kd)^{n-1/2},
		\label{BeverlooEq1}
	\end{equation}
	
	\noindent where $n$ is the spatial dimension, i.e., $n=2$ or 3 for two or three dimensions, respectively. $d$ denotes the grain diameter, $g^*$ denotes the effective acceleration due to gravity. $C$ represents a dimensional fitting coefficient that may be associated with factors such as the friction coefficient and bulk density. $k$ is believed to result from the empty$-$annulus effect related to the geometrical shape of the grain and the outlet. Within this argument, the grains are strongly crowed in the hopper and the lasting collisions between the grains dissipate the gravity energy. This high packing density leads to the occurrence of Jassen effect\cite{Janssen1895}. 
	A hypothesis is thus proposed regarding existence of a half$-$circular shape of contact force around the outlet, named as {\it free fall arch} (FFA). The initial velocities of the grains are negligible and they freely accelerate downward under the self gravities below the FFA. In the past decades, the expression of Eq.\ref{BeverlooEq1} has been verified for various situations, such as different kind of grains in size and shape, different outlet geometrical configuration. 
	\cite{HuangPT2021,NeddermanCES1982,Zuriguel2012PRL108P248001,AguirrePRL2010, ZhengPT345P676Y2019,Peng2021POF33P043313, Maiti2017POF29P103303,ArteagaCES1990,HumbyCES1998,BenyaminePRE2014,Huang2018CPB,HuangPT2022}. 
	Some modifications to the coefficients of $C$ and $k$ have been proposed to adapt the change of flow situations. However, the physical mechanism of Beverloo law is retained, in which the basic hypothesis of the FFA is considered to dominate the flow properties. 
	Recently, a controversy was recently put forward by Janda et al. and their experimental and numerical results are not supported by the arguments of FFA, especially for the case of small outlet.\cite{Zuriguel2012PRL108P248001} A self$-$similarity in dense flow state for the packing density and grain velocity are first discovered. A new expression is proposed as the following:
	\begin{equation}
		Q=C'{\sqrt {g^*}}(1-{\alpha_1}e^{-R/{\alpha_2}})R^{3/2},
		\label{JandaEq1}
	\end{equation}
	
	\noindent where $C$ is a parameter related to the grain diameter $d$ and packing density at the outlet, $\alpha_1$ and $\alpha_2$ are fitting coefficients affected by the packing density at the center of the outlet, and $R=D/2$ is the half of exit size $D$. This local self$-$similar behavior around the outlet are examined under different conditions, such as the bidispersed flow and eccentrical$-$outlet hopper.\cite{BenyaminePRE2014,ZhouPRE2015,BhatejaPOF2022} These studies have demonstrated that changes in grain properties and geometrical configuration of the hopper do not affect the local self$-$similar behavior. 
	
	In the Beverloo and Janda arguments, the local flow properties around the outlet have attracted much attention. However, the global flow properties at the positions far from the outlet, i.e., mass and funnel flow patterns, and the transition between them, have been found to significantly affect on the dense flow rate. \cite{HuangPT2022,Nedderman1979PT22P243,RobertsCES2002,WojcikPT2012,ZhangPT2018,JiShunyingPT2019,JiShunyingSM2020,KalyanPT2023} 
	In the mass flow pattern, all the grains flow simultaneously in the hopper having uniform and slow velocities, whereas in the funnel flow pattern, grains exhibit higher velocities at the hopper center than those near the side walls. The former is typically observed in regions far from the outlet, whereas the latter is found near the outlet. The transition height is a critical value affecting the dense flow rate. In the study of Ji {\it et al.}, an external pressure applied to the top of the hopper causes the transition height to move closer to the outlet, leading to an increase in the flow rate.\cite{JiShunyingPT2019,JiShunyingSM2020} In the work of Kalyan {\it et al.}, it was noted that flow with bulky$-$shaped grains does not transition to the mass flow pattern; instead, the funnel flow pattern dominates the entire flow within the hopper.\cite{KalyanPT2023} This flow pattern results in a higher flow rate compared to flows comprising elongated or angular grains. 
	Our previous research indicated that a well$-$designed mixed binary flow containing different$-$sized grains can achieve the maximum dense flow rate.\cite{HuangPT2022} The findings suggest that an earlier transition from the mass flow pattern to the funnel flow pattern can enhance grain velocity and consequently increase the flow rate. 
	Other factors that have been identified to affect the dense flow rate include outlet geometry and the presence of obstacles in the hopper.\cite{ZuriguelPRL2011,ZuriguelPRF2017,EndoPRF2017,ZuriguelPRL2023} However, the hopper width stands out as another critical factor whose influence remains to be fully understood. 
	
	In this study, both experiments and numerical simulations are conducted to investigate the impact of hopper width on dense granular flow in a two$-$dimensional hopper. The experimental setup and corresponding simulation model are detailed in Section \ref{SimMod}. In Section \ref{Results}, we first explore the influence of hopper width on the dense granular flow rate, followed by an examination of the relationship between the dense flow rate and outlet size using the Beverloo and Janda laws. Subsequently, we analyze the flow properties at both local and global scales. We revisit the self$-$similarities concerning packing density and grain velocity around the outlet and introduce the global similarity regarding the transition between the mass flow pattern and the funnel flow pattern. Finally, our main results are summarized in Section \ref{Concl}.
	
	\section{Experiment setup and simulation model}
	\label{SimMod}
	
\begin{figure}[htbp]
		\centering
		\includegraphics[width=7 cm,trim=100 230 100 230,clip]{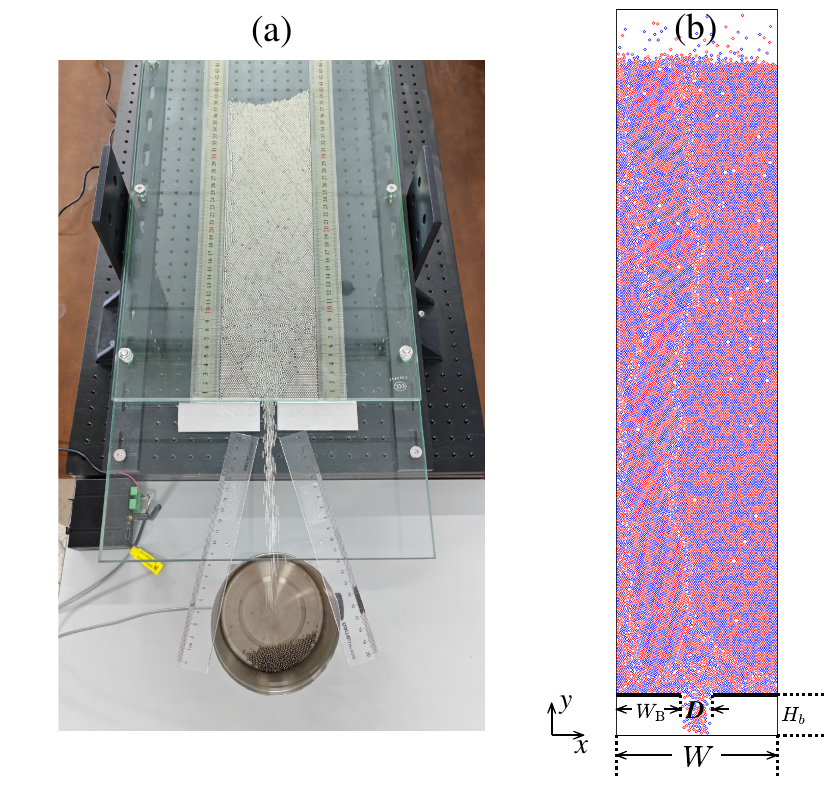}
		\caption{(Color online). Snapshots of (a) experiment setup and (b) simulation system.}
		\label{fig:FigModel}
	\end{figure}

	The experiments are carried out in a quasi two$-$dimensional(2D) rectangular hopper with an inclination angle of $15^{\circ}$ as shown in Fig. \ref{fig:FigModel}(a). The hopper is built with two glass plates on the bottom and the top, which are separated by two steel stripes that also act as lateral walls. The gap between the glass plates is $2.2$ mm to allow for a quasi single$-$layer flow of steel grains with a diameter of $d=2\pm0.01~{\rm mm}$.
	
	A central outlet is formed by two flat aluminum baffles at the bottom of the hopper. The steel stripes and aluminum baffles are movable to adjust the hopper width $W$ and the outlet size $D$. The hopper is empty at the beginning time and the outlet is closed by placing an baffle. The steel grains are poured randomly into the hopper and an initial piling height is $H_{\rm p} \approx 400~{\rm mm}$. Then, the baffle is removed and the flow is triggered. The discharged grains are collected by a container which is placed on force sensor (LH$-$Z05A) with a precision of 1 g at 50 Hz. The temporal evolution of the mass of the grains is recorded and the flow rate is determined by calculating the slope of the mass. Each experiment is repeated five times to get the average value.
	
	In the simulations, the same quasi$-$2D hopper is used with a height of $H=450~{\rm mm}$ and a thickness $T=2~{\rm mm}$ as shown in Fig. \ref{fig:FigModel}(b). Two flat baffles are placed at the height $H_{\rm b}=25.0~{\rm mm}$ to form the outlet with size $D$. The widths of the baffle $W_{\rm B}$ can be changed. Discrete element method is used to investigate the granular flow and the grain motion is described using Newton\('\)s equations, as in our previous works\cite{HouPRL2003,Huang2006PRE,Huang2011PLA,HuangPT2021,Huang2018CPB,HuangPT2022}.
	The effective acceleration due to gravity is set as $g^{*}=\sin15^{\circ}~{\times}~9.8~{\rm m/s^{2}}$ to keep consistent with the experiment. The positions and velocities of the grains at each simulation time step are updated by using the Verlet algorithm. 
	The translational motion in the hopper plane and the rotational motion perpendicular to the hopper plane are both considered. The contact interaction between two grains is calculated in the normal and tangential directions. The normal force at the contact point is modeled by the Cundall$-$Strack form \cite{Cundall1979,Schafer1996}:
	\begin{equation}
		F_{n}={\frac{4}{3}}E^{*}{\sqrt{R^{*}}}{\delta_{n}}^{3/2}
		-2.0{\sqrt{\frac{5}{6}}}{\beta}{\sqrt{S_{n}m^{*}}}V_{n}.
		\label{CundallFn}
	\end{equation}
	
	The tangential component is considered as the minor tangential force with a memory effect and the dynamic frictional force:
	\begin{equation}
		F_{\tau}=-\min({S_{\tau} {\delta}_{\tau}}-2.0{\sqrt{\frac{5}{6}}}{\beta}{\sqrt{S_{\tau}m^{*}}}V_{\tau} {{\mu} F_{n}}){\rm{sign}}({\delta}_{\tau}).
		\label{CundallFt}
	\end{equation}
	
	In Eqs. (\ref{CundallFn}) and (\ref{CundallFt}), $n$ and $\tau$ respectively denote the normal and tangential directions at the contact point, and $\delta_{n}$ and $\delta_{\tau}$ denote the normal and tangential displacements since time $t_0$ at which contact is initially established. The calculation details as follows:
	\begin{equation*}
		\beta={\frac{{\rm ln}e}{{\sqrt{{\rm ln}^{2}e}+{\pi}^{2}}}},~
		{\frac{1}{m^{*}}}={\frac{1}{m_{i}}}+{\frac{1}{m_{j}}}, \\
	\end{equation*}
	\begin{equation*}
		S_{n}=2E^{*}{\sqrt{R^{*}{\delta_{n}}}},~
		S_{\tau}=8E^{*}{\sqrt{R^{*}{\delta_{n}}}},\\
	\end{equation*}
	\begin{equation*}
		E^{*}={\frac{1-{\nu_{i}^{2}}}{E_{i}}}+{\frac{1-{\nu_{j}^{2}}}{E_{j}}},~
		{\frac{1}{R^{*}}}={\frac{1}{R_{i}}}+{\frac{1}{R_{j}}}, \\
	\end{equation*}
	
	\noindent where $e$ is the coefficient of restitution. The quantities $m_i$ and $m_j$ are the masses of grains $i$ and $j$ making contact, respectively, and $S_{n}$ and $S_{\tau}$ characterize the normal and tangential stiffness of the grains. $E$ and ${\nu}$ denote the Young\('\)s modulus and Poisson\('\)s ratio, respectively. In our simulations, the friction coefficient $\mu$ is fixed, and a collision between a grain and a wall is treated as a grain$-$grain collision, except that the wall has infinite mass and diameter. Table I lists the values of the material parameters of the grains.
	\begin{center}
		{\bf{Table I}} Grain parameters \\
		\begin{tabular} { p{5.8cm} p{1.45cm} p{0.95cm} }
			\hline
			Quantity & Symbol   & Value \\
			\hline
			Diameter of grain [mm] & \emph {$~~~d$} & 2.0 \\
			Density [$\rm {10^{3}kg/m}^{3}$] & \emph {$~~~\rho$} & 7.8 \\
			Young\('\)s modulus [GPa] & \emph {~~~E} & 1.0 \\
			Poisson ratio & $~~~\nu $ & 0.3 \\
			Friction coefficient & $~~~\mu$ & 0.5 \\
			Simulation time-step [s] & $~~~dt$ & $10^{-6}$ \\
			\hline
		\end{tabular}
		\label{TableGrainPara}
	\end{center}
	
	The process of simulation is similar to that used in the experiments. The outlet is closed at the beginning of simulation. Grains are generated randomly at the top of the hopper and fall down under gravity. The initial piling height of grains is set to $H_{\rm p} \approx 400~{\rm mm}$. The grains start to flow when the outlet is opened. Periodical condition is used, in which the grains flowing out the hopper reenter the hopper from the top. Thus a fixed piling height is kept and a stable dense granular flow can be obtained after two seconds in the simulation. The position and velocity of each grain and the interaction between the grains are recorded for the calculation.
	
	\begin{figure}[htbp]
		\centering
		\includegraphics[width=7 cm,trim=100 160 110 165,clip]{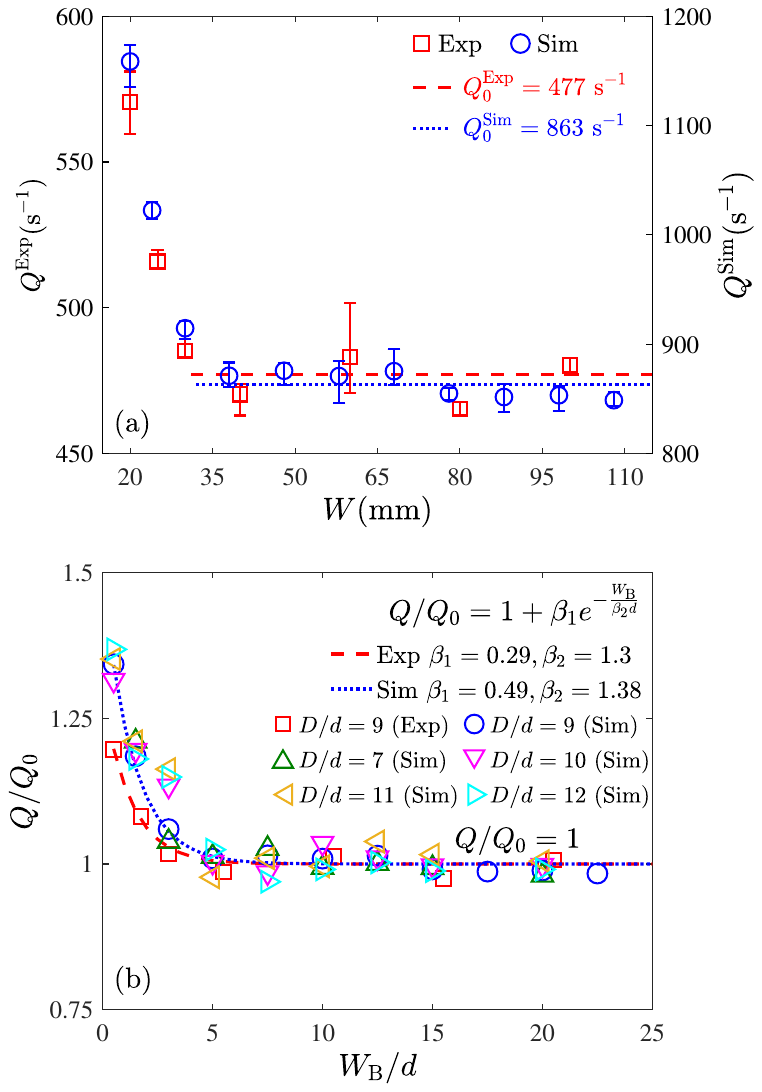}
		\caption{(Color online). (a) Outflow rate as a function of hopper width for experiment and  simulation results. The outlet size is fixed at $D=18~{\rm mm}$. The dash and the dotted lines donate the mean flow rate of large hopper widths. (b) Outflow rate normalized by the mean flow rate of large hopper widths as a function of scaled baffle length. The dash and the dotted lines donate the fitted results for $D/d=9$ by using Eq.(\ref{EqQoutWB}). The other outlet sizes, $D/d=7,10,11$ and 12, are used in the simulations and donated by different types of symbols.}
		\label{fig:FigQoutWidth}
	\end{figure}
	
	\section{Results and discussion}
	\label{Results}
	
	The experiment and simulation results of flow rate as a function of hopper width are plotted in Fig. \ref{fig:FigQoutWidth}(a). The experiment results  exhibit similar flow properties to the simulations. When the hopper width is sufficiently large, the flow rate remains at a constant value, as indicated by dash lines, i.e., $Q_{0}^{\rm Exp}=477~{{\rm s}^{-1}}$ and $Q_{0}^{\rm Sim}=863~{{\rm s}^{-1}}$. However, the presence of the sidewalls begins to influence the flow rate for smaller hopper widths. A decrease in hopper width leads to a significant increase in the flow rate. Moreover, the simulation results are noticeably higher than the experimental results, which may be attributed to the friction between the grains and the two glass plates at the bottom and top in the experiments. 
	
	In Fig. \ref{fig:FigQoutWidth}(b), an exponential scaling law  is obtained describing the relation between the flow rate and the baffle length for outlet size $D/d=9$ as the following:
	
	\begin{equation}
		{Q/Q_0}=1+{\beta_1}e^{-\frac{W_{\rm B}}{{\beta_2}d}}
		\label{EqQoutWB}
	\end{equation}
	
	\noindent where $\beta_1$ and $\beta_2$ are fitting parameters.
	We can see that the fitting parameter $\beta_1$, 0.29 for the experiment and 0.49 for the simulation, indicates the maximum increase of flow rate corresponding to the baffle with an infinite small length compared to that with a large baffle length. The other fitting parameter $\beta_2$, 1.3 for the experiment and 1.38 for the simulation, represents the critical baffle length below which the sidewall starts to significantly influence on the flow rate. Additionally, several other outlet sizes, i.e., $D/d=7,10,11$ and 12, are also used in the simulations, and all the results collapse well with the fitted curve as shown in Fig. \ref{fig:FigQoutWidth}(b). This good agreement suggests that the flows with different baffle lengths share some inherent common kinematic characteristics. Therefore, these characteristics the focus of the next study. 
	
	\begin{figure}[htbp]
		\centering
		\includegraphics[width=7 cm,trim=120 290 125 290,clip]{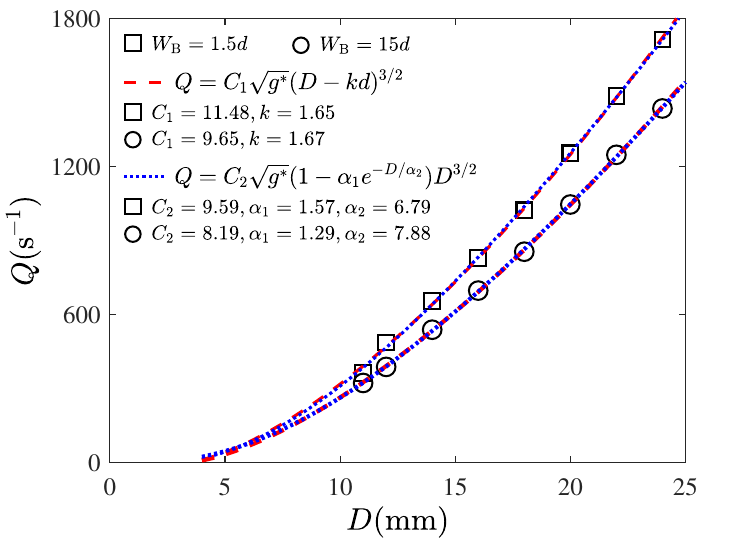}
		\caption{(Color online). Simulated results of flow rate as a function of outlet size for different baffle lengths $W_{\rm B}=1.5$ and $15d$. The red dash curves and blue dotted curves are obtained by using the Beverloo and the Janda equations, Eqs.(\ref{BeverlooEq1}) and (\ref{JandaEq1}), respectively.}
		\label{fig:FigQoutD}
	\end{figure}
	
	The first question is whether the classical theories proposed by Beverloo and Janda et al. can describe the relation between the flow rate and the outlet size. Two baffles with lengths of $W_{\rm B}/d=1.5$ and $15$ are used, as shown in Fig.\ref{fig:FigQoutD}. The simulation results agree well theoretical predictions, indicating that the flow rate increases proportionally to the outlet size in the power of $D^{3/2}$. This characteristic suggests that the local mechanical properties around the outlet significantly dominate the relationship between the flow rate and the outlet size. As a result, the obtained fitting parameters shown in Fig.\ref{fig:FigQoutD} are generally consistent with the original hypothesis. Large values of $C_1=11.48$ and $C_2=9.59$ are obtained for a narrow hopper with a baffle length of $W_{\rm B}=1.5d$ compared to smaller values of $C_1=9.65$ and $C_2=8.19$ for wide hopper with baffle length $W_{\rm B}=15d$, respectively. However, there are some discrepancies. In the Beverloo case, the effective outlet size does not change significantly, i.e., $k=1.65$ and $1.67$ for $W_{\rm B}=1.5d$ and $15d$, respectively. This implies that grain velocity has little influence on the flow rate. 
	In Janda case, the packing density appears to have a minor change on the flow rate because the variation in silo width has little effect on $\alpha_1$ and $\alpha_2$. 
	
	\begin{figure*}[htbp]
		\centering
		\includegraphics[width=0.7\textwidth,trim=120 210 115 270,clip]{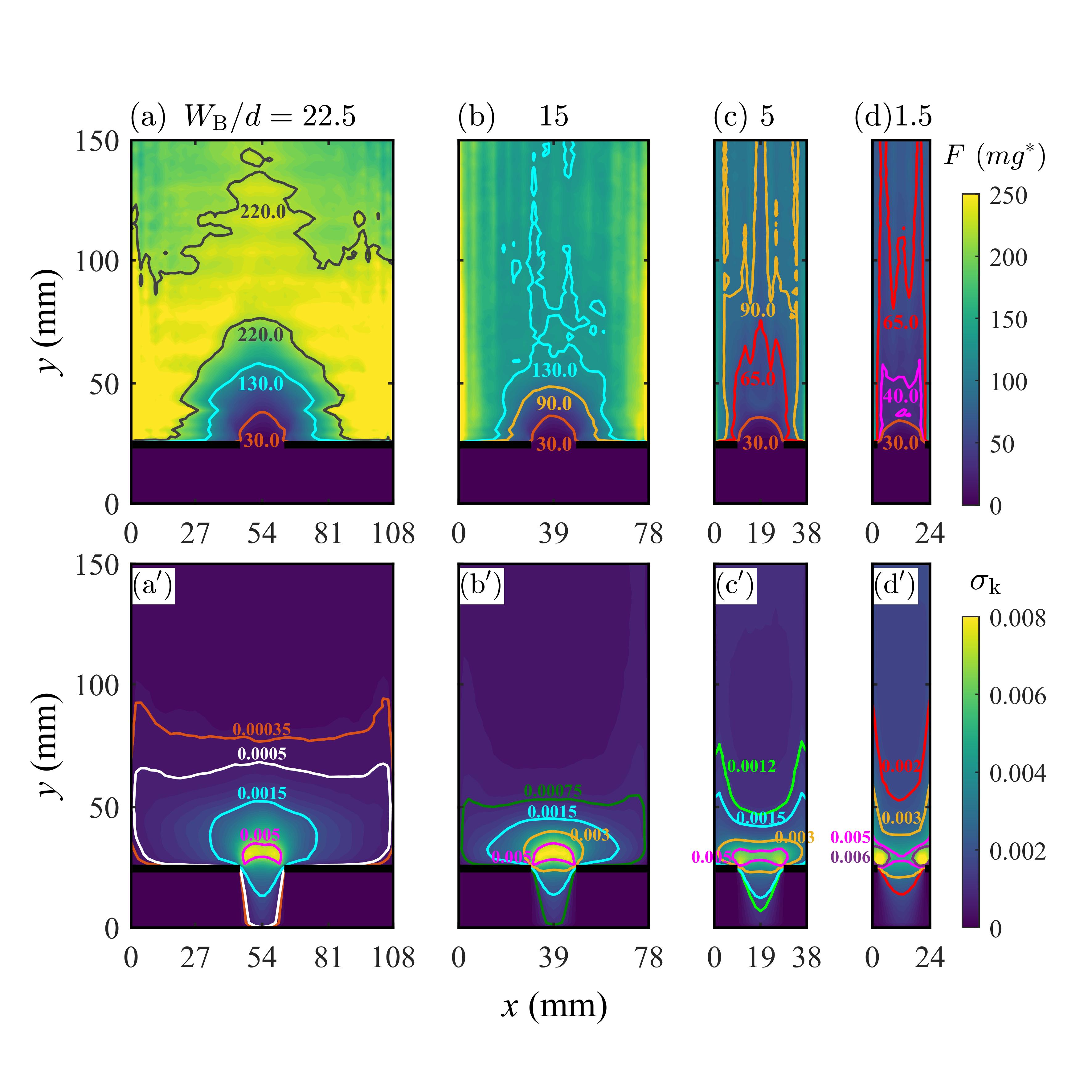}
		\caption{(Color online). Spatial distribution of ({\rm a})-({\rm d}) contact force between grains and (a\('\))-(d\('\)) the kinetic force. Four baffle lengths are used, i.e., (a)(a\('\))$W_{\rm B}/d=22.5$, (b)(b\('\))$W_{\rm B}/d=15$, (c)(c\('\))$W_{\rm B}/d=5$, (d)(d\('\))$W_{\rm B}/d=1.5$.} 
		\label{fig:FigFSigmakDist}
	\end{figure*}
	
	Fig.\ref{fig:FigFSigmakDist}(a) plots the spatial distribution of contact force between grains in a flow with a large baffle length of $W_{\rm B}/d=22.5$. In the regions far from the outlet, the contact forces between grains are evenly distributed, with the contour lines appearing nearly horizontal. As the grains continues to flow downward, a large arch$-$shaped structure with strong mutual contact forces becomes prominent throughout the entire hopper in the regions labeled as $220.0$. This arch$-$shaped structure is maintained until the grains reach the region just above the outlet. The magnitude of the contact force gradually decreases, as indicated by labels $130$ and $30$.
	When the baffle length is reduced, i.e., $W_{\rm B}/d=15$, as shown in Fig.\ref{fig:FigFSigmakDist}(b), a series of arch$-$shaped contact forces occur in the region just above the outlet, labeled as $90$ and $30$. However, the adjacent arch$-$shape structure begins to deform, as indicated by $130$.
	When the baffle length continues to decrease, i.e., $W_{\rm B}/d=5$, only a localized arch$-$shape structure is still observed labeled as $30$, as shown in Fig.\ref{fig:FigFSigmakDist}(c). Significant deformation has occurred in the arch$-$shaped structure for the contact forces of $65$ and $90$s. This deformation becomes more pronounced for the contact forces of $65$ and $40$ when the baffle length is $W_{\rm B}/d=1.5$, as shown in Fig.\ref{fig:FigFSigmakDist}(d). Only the arch$-$shaped structure labeled as $30$ is locally maintained just above the outlet.
	The presence of the arch$-$shape structure in the contact force above the outlet provides substantial evidence for the hypothesis of FFA. On the other hand, the deformation of the contact force structure far from the outlet, especially for small baffle length, indicates the occurrence of the transition of flow states within the hopper.
	
	\begin{figure*}[htbp]
		\centering
		\includegraphics[width=0.7\textwidth,trim=120 210 125 270,clip]{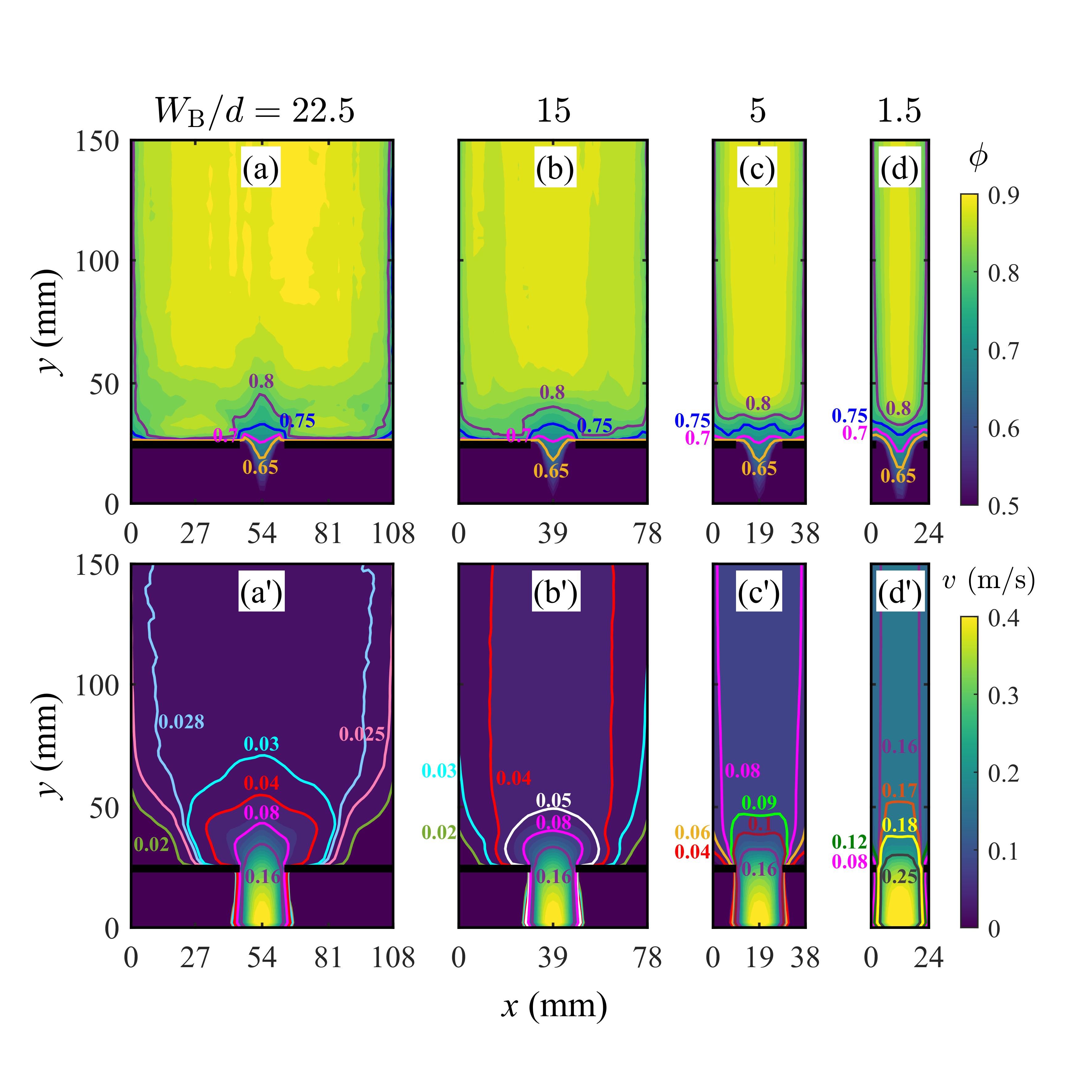}
		\caption{(Color online). Spatial distribution of (a)-(d) packing density and (a\('\))-(d\('\)) grain velocity for different baffle lengths. The outlet size is $D=18~{\rm mm}$. The solid lines of different colors indicate the contours of different values.}
		\label{fig:FigPackVelDist}
	\end{figure*} 
	
	Based on Rubio$-$Largo's arguments\cite{ZuriguelPRL2015}, the spatial distributions of the kinetic force are plotted in Fig.\ref{fig:FigFSigmakDist}(a\('\))(b\('\))(c\('\))(d\('\)). When the baffle length is large, such as $W_{\rm B}/d=22.5$, the maximum kinetic force, such as $0.005$, is localized. It just crosses the top of the outlet, forming an arch$-$shape structure as shown in Fig.\ref{fig:FigFSigmakDist}(a\('\)). Compared to the contact force shown in Fig.\ref{fig:FigFSigmakDist}(a), the kinetic force decreases along the  direction opposite to that of granular flow, indicated  $0.0015,0005$ and $0.00035$. More importantly, the contour of the equivalent kinetic force in the regions far from the outlet is extended, crossing the entire width of the hopper in the transverse direction. Their shapes also change from a hemispherical form to a more flat form appearance. 
	In Fig.\ref{fig:FigFSigmakDist}(b\('\)), a similar evolution of the kinetic force is observed when $W_{\rm B}/d=15$. Larger kinetic forces are localized and form an arch$-$shape structure above the outlet. As the flow moves away from the outlet along in opposition to the direction of granular flow, the kinetic force decreases, marked by labels $0.003,0.0015$ and $0.00075$. Simultaneously, as the kinetic force extends across the entire width of the hopper, the height of the corresponding enclosed structure diminishes.
	When the baffle length is further decreased, i.e., $W_{\rm B}/d=5$, a significant change in the evolution of the kinetic force occurs, as shown in Fig.\ref{fig:FigFSigmakDist}(c'). Although the larger kinetic forces still occur around the outlet, they now form a dumbbell$-$shaped distribution, marked with a label of $0.0005$. This distribution extends across the entire hopper while decreasing in magnitude opposite to the direction of the flow, as indicated by label $0.0003$. When the flow is far from the outlet, a U$-$shaped structure of the kinetic force is observed, corresponding to labels $0.0015$ and $0.0012$. 
	In Fig.\ref{fig:FigFSigmakDist}(d\('\)), a small baffle length $W_{\rm B}/d=1.5$ is used. The U$-$shaped structure of the kinetic force, marked by labels $0.003$ and $0.002$, is maintained in regions far from the outlet. However, for the contour lines of the kinetic force around the outlet, only two hammers in the dumbbell shape remain, located at the sides of the outlet, and labeled as $0.006$. For the contour line of $0.005$, it comprises an upward triangle and a downward triangle.
	
	\begin{figure*}[htbp]
		\centering
		\includegraphics[width=0.95\textwidth,trim=15 260 25 260,clip,angle=0]{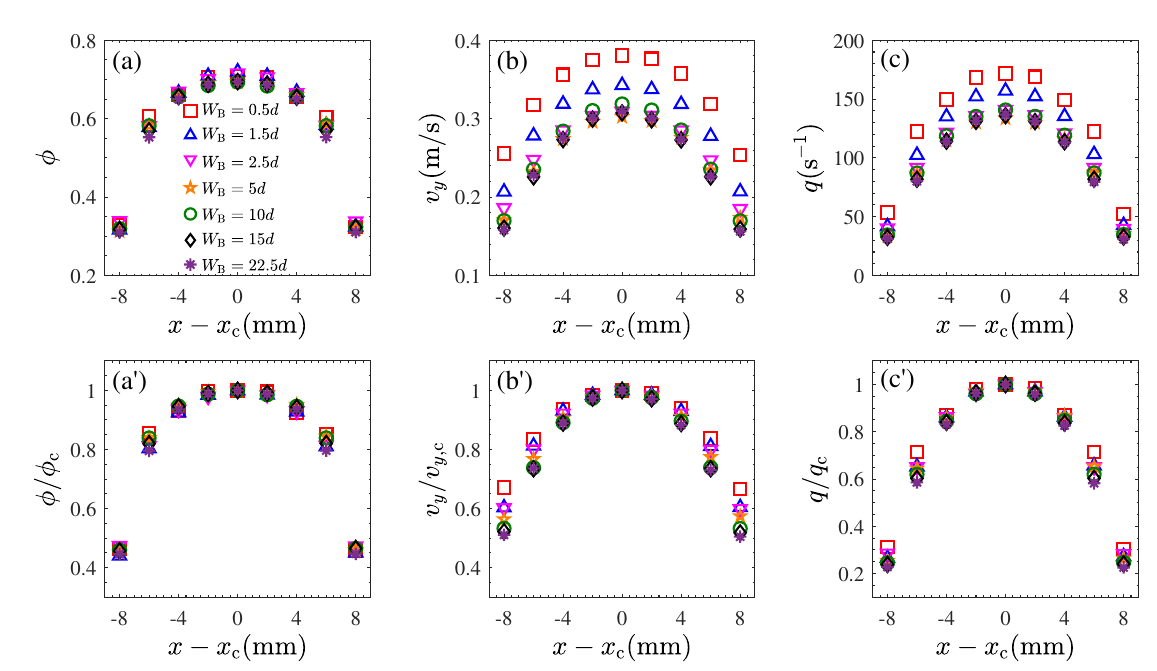}
		\caption{(Color online). Self$-$similar horizontal profiles of (a) vertical velocity, (b) packing density, and (c) flow rate, and corresponding normalized profiles of (a\('\)) vertical velocity, (b\('\)) packing density, and (c\('\)) flow rate for different baffle widths. $x_{\rm c}$ is the location of outlet center. The outlet size is $D=18~{\rm mm}$.}
		\label{fig:FigPackVyQExit}
	\end{figure*}
	
	The results of the contact force and kinetic force strongly support the arguments proposed by Beverloo's and Janda's formula regarding the local properties around the outlet. Moreover, variations in baffle length have a significant impact on the contact force and kinetic force in regions around the outlet. This leads to two distinct flow states: mass flow far from the outlet and funnel flow near the outlet. In conditions of large baffle length, the mass flow pattern is observed over a large$-$scale region due to the presence of a significant arch$-$shaped contact force. This arch$-$shaped structure provides robust support for grains located far from the outlet, facilitating their movement in a mass flow pattern. Conversely, when a U$-$shaped structure of the kinetic force emerges, it indicates the presence of funnel flow. As the baffle length decreases, the transition point from mass flow to funnel flow occurs further away from the outlet. This shift in flow dynamics plays a crucial role in the sudden increase in flow rate. Additionally, packing density and grain velocity are two key factors that determine the flow rate. Next, we will examine how these quantities change in their spatial distribution as the hopper width varies.
	
	The spatial distribution of packing density is first analyzed for the baffle length of $W_{\rm B}=22.5d$ shown in Fig.\ref{fig:FigPackVelDist}(a). For the flow far from the outlet, the mass flow pattern occurs, where the grains are highly crowded together, $\phi>0.8$. This high packing density leads to a strong contact force between the grains. The packing density starts to decrease when the grains reach the region above the outlet forming an arch$-$shaped structure, while still maintaining a high value, $\phi=0.8$ above the outlet. As the grains continue to flow, approaching the outlet, the packing density decrease further and the arch$-$like structure flattens, indicated by $\phi=0.75$. This suggests that self$-$gravity has started to dominate the flow, pushing the grains into a funnel flow pattern. The packing density continuously decreases, marked by $\phi=0.7$ and $0.65$, until the grains flow across the outlet. These flow properties in packing density are observed again for the baffle length of $W_{\rm B}/d=15$ as shown in Fig.\ref{fig:FigPackVelDist}(b). The grains first flow downward in a mass flow pattern with a high packing density, indicated by $\phi=0.8$. Around the outlet regions, a funnel flow pattern occurs in which the packing density decreases gradually, marked by labels $0.75,0.7$ and $0.65$.
	When the baffle length is $W_{\rm B}/d=5$, a U$-$shaped structure of packing density labeled as $0.8,0.75$ and $0.7$ is observed, as shown Fig.\ref{fig:FigPackVelDist}(c). This characteristic indicates that the transition between the mass flow pattern and the funnel flow pattern has occurred. Compared with the results in Figs.\ref{fig:FigPackVelDist}(a)(b), the position of the transition between the mass flow pattern and the funnel flow pattern has further shifted upward. This shift becomes more significant when a small baffle length is used $W_{\rm B}/d=1.5d$ shown in Fig.\ref{fig:FigPackVelDist}(d). 
	A series of U$-$shaped structure of packing density with a convex bottom occur, labeled as $0.8,0.75$ and $0.7$. Furthermore, this convex U$-$shaped structure extends across the outlet, marked by the label $0.65$.
	
	Similarly, two different spatial distributions of grain velocity are reproduced, as shown in Figs.\ref{fig:FigPackVelDist}(a\('\))(b\('\)) (c\('\))(d\('\)). The first one is an open shape, located in regions far from the outlet. One side of the contour line is located adjacent the baffle, while the other side extends towards the hopper wall or the upper side of the hopper. The second flow state is located around the outlet and forms a closed loop. When the size of the outlet decreases, the occupied region of the former expands, while that of the latter shrinks. Additionally, the shape of the latter changes from a half circle to a reversed U$-$shape. Consequently,the mass and funnel flow patterns occurs in the occupied regions having the open and closed shapes of grain velocity. Furthermore, reducing the baffle length leads to an increase in the height of the transition between the mass and funnel flow patterns.
	
	Given that the hopper width significantly influences the spatial distribution flow dynamics in both local and global scales, it is essential to perform a comprehensive quantitative analysis of packing density and grain velocity. Let us first present the horizontal profile of packing density $\phi$ at the outlet with a size of $D=18{~\rm mm}$ for different baffle widths, as shown in Fig.\ref{fig:FigPackVyQExit}(a). It is evident that a symmetric profile is observed independently of the baffle length. The self$-$similar characteristic is found when considering the corresponding normalized case using the maximum packing density $\phi_{\rm c}$ at the outlet center, as shown in Fig.\ref{fig:FigPackVyQExit}(a\('\)). 
	The horizontal profile of the vertical grain velocity $v_{y,\rm c}$ is plotted in Fig.\ref{fig:FigPackVyQExit}(b). A decrease of the baffle length clearly increases the grain velocity in the vertical direction $v_{y,\rm c}$m at the outlet. Once again, the self$-$similar profile is observed in the corresponding normalized case using the maximum grain velocity in the vertical direction $v_{y,\rm c}$ at the outlet center shown in Fig.\ref{fig:FigPackVyQExit}(b\('\)). 
	These self$-$similar properties of packing density and grain velocity are also maintained in the flow rate profiles, as shown in Figs.\ref{fig:FigPackVyQExit}(c)(c\('\)). 
	Furthermore, although a decrease in baffle length results in an increase in flow rate, the normalized results for different baffle widths collapse remarkably well together. These results provide strong evidence for the presence of self$-$similarity in the packing density, grain velocity, and flow rate at the local region around the outlet, as suggested by Janda et al.
	
	\begin{figure}[htbp]
		\centering
		\includegraphics[width=7 cm, trim=115 280 115 250,clip]{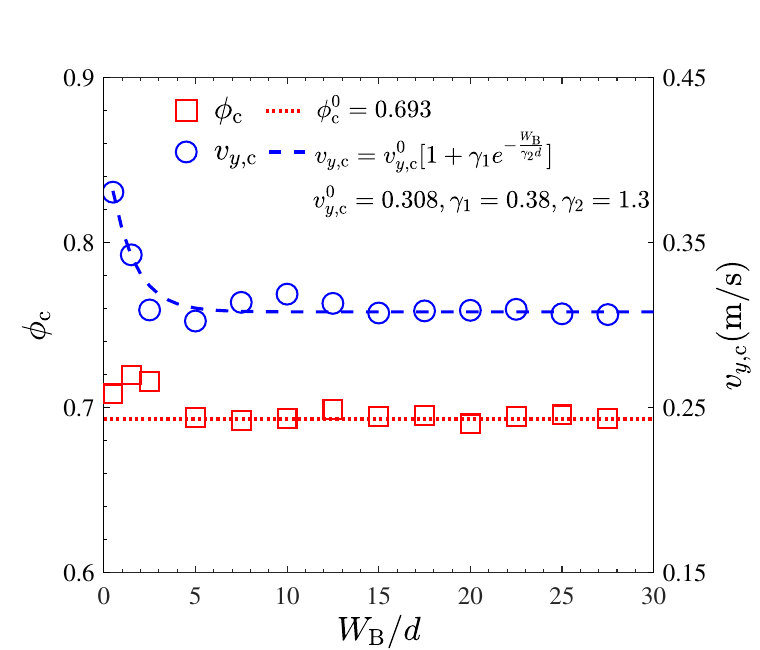}
		\caption{(Color online). Packing density (squares) and mean vertical velocity (circles) at the center of the exit as a function of baffle length. The width of the outlet is $D=18~{\rm mm}$. The dashed curve is obtained by using Eq.(\ref{eq:EqVyEC}).}
		\label{fig:FigPackVyExitC}
	\end{figure}
	
	Now, we consider the influence of the baffle length on the packing density and the vertical grain velocity at the outlet center, as shown in Fig.\ref{fig:FigPackVyExitC}. It can be observed that the change in hopper width has little impact on the packing density, which remains constant at $\phi_{\rm c}=0.693$, although a slight increase occurs for small hopper widths. Similarly, the grain velocity in the vertical direction remains constant at $v_{y,\rm c}^0=0.306~\rm m/s$ for large hopper widths. However, for small hopper widths, a sudden increase in grain velocity occurs, similar to the sudden increase observed in flow rate. Therefore, it is reasonable to use a similar equation to describe the relationship between grain velocity and hopper width.
	\begin{equation}
		v_{y,\rm c}=v_{y,\rm c}^0[1+{\gamma_1}e^{-\frac{W_{\rm B}}{{\gamma_2}d}}]
		\label{eq:EqVyEC}
	\end{equation}
	
	The simulation results of the grain velocity show good agreement with Eq. \ref{eq:EqVyEC}, where $\gamma_1=0.36$ and $\gamma_2=1.29$. Compared with the results shown in Fig. \ref{fig:FigQoutWidth}, the similarity between the flow rate and the grain velocity indicates that the grain velocity plays a more significant role in determining the flow rate than the packing fraction in the local region around the outlet.
	
	\begin{figure*}[htbp]
		\centering
		\includegraphics[width=0.95\textwidth,trim=20 230 20 220,clip]{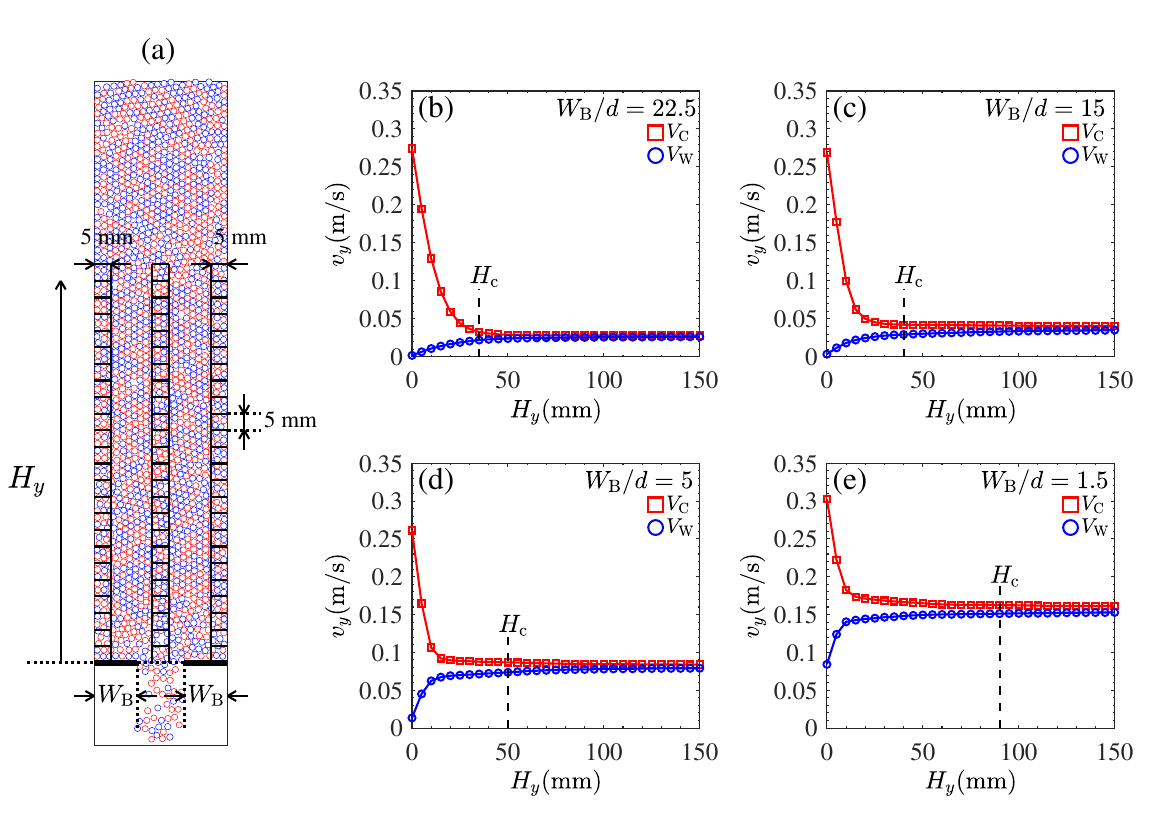}
		\caption{(Color online). (a) Snapshot of extraction area for granular vertical velocity. Vertical grain velocity of selected areas as a function of the packing height for baffle length (b) $W_{\rm B}/d=22.5$, (c) $W_{\rm B}/d=15$, (d) $W_{\rm B}/d=5$ and (e) $W_{\rm B}/d=1.5$. The width of the outlet is $D=18~{\rm mm}$.}
		\label{fig:FigVyH}
	\end{figure*}
	
	\begin{figure}[htbp]
		\centering
		\includegraphics[width=7 cm, trim=100 260 120 240,clip]{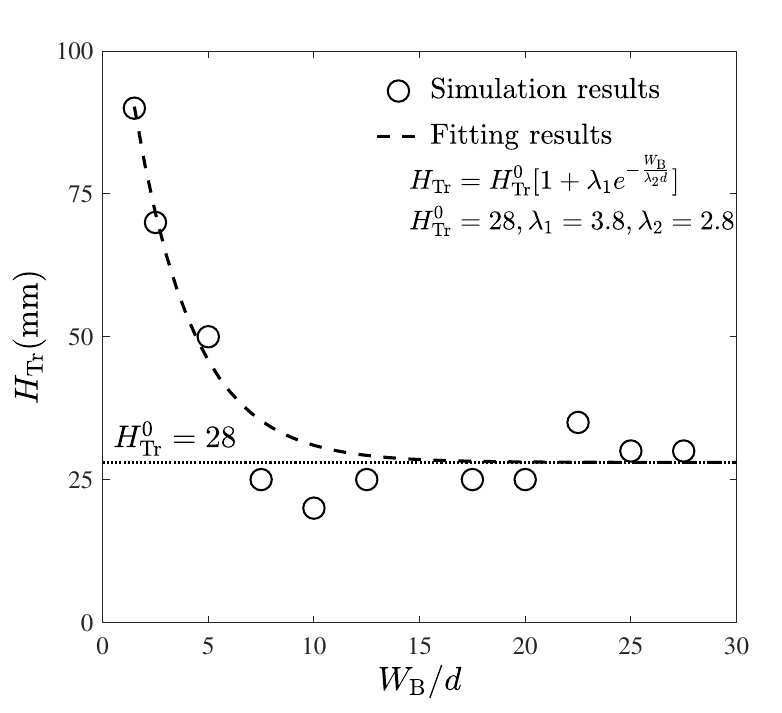}
		\caption{Relationship between the transition height and the baffle length. Circles are the simulation results. Dashed line is the fitting result by using the equation $H_{\rm Tr}=H_{\rm Tr}^0[1+{\lambda_1}e^{-\frac{W_{\rm B}}{{\lambda_2}d}}]$. Dotted line indicates the critical transition height ${H_{\rm Tr}^0}=28$.}
		\label{fig:FigHcWB}
	\end{figure}
	
	Since the relationship between the flow rate and grain velocity has exhibited inherent self$-$similarity at the outlet, we now investigate their dependence on a global scale. To analyze the transition between the mass flow pattern and the funnel flow pattern, we employ the same method as previously reported in our work\cite{HuangPT2022}. In this analysis, three narrow rectangular regions are subdivided into numerous squares with edge length $5{\rm mm}$, as shown Fig. \ref{fig:FigVyH}(a).
	The dependence of the vertical grain velocity $v_{y}$ on the granular packing height $H_{y}$ is plotted in Figs. \ref{fig:FigVyH}(b)$-$\ref{fig:FigVyH}(d) for $W_{\rm B}/d=1.5,5,15$ and $22.5$, respectively. Given the symmetry, the results at two sidewalls are averaged. The transition height $H_{\rm Tr}$ is introduced to describe the position of the transition between the mass flow pattern and the funnel flow pattern. $H_{\rm Tr}$ is obtained when $\Delta V_y = V_{\rm C} - V_{\rm W} < \Delta V_{y,\rm max}{\times}5\%$, where $V_{\rm C}$ and $V_{\rm W}$ are the grain velocities in the vertical direction at the positions of center and sidewall.
	When $H_{y}>H_{\rm Tr}$, the flow has a uniform grain velocity in the vertical direction, indicating the mass$-$flow state. When $H_{y}<H_{\rm Tr}$, the funnel$-$flow state occurs, where grains at the center of the hopper have higher velocities than those near the sidewalls. A reduction in hopper width tends to increase the transition height $H_{\rm Tr}$. Therefore, Fig. \ref{fig:FigHcWB} plots the relationship between the transition height and the baffle length. A similar asymptotic form is observed as in Fig. \ref{fig:FigQoutWidth} and Fig. \ref{fig:FigPackVyExitC}. For sufficiently large baffle lengths, a stable characteristic transition height of ${H_{\rm Tr}^0}=28$ is obtained. However, for smaller baffle lengths, reducing the baffle length leads to a significant jump in the transition height.
	The existence of the characteristic transition height indicates that considering only local flow properties is insufficient for modeling dense flow rates. Consequently, the transition between the mass flow pattern and the funnel flow pattern on a global scale results in an exponential increase in the dense flow rate.
	
	\section{Conclusions}
	\label{Concl}
	
	In this work, both experiments and discrete element simulations are conducted to explore the influence of hopper width on dense granular flow in a 2D hopper. The flow rate remains constant when the hopper width is large, whereas a significant increase in flow rate is observed as the hopper width decreases. Additionally, the relationship between the flow rate and baffle length can be described by an exponential scaling law over a wide range of outlet sizes, indicating the self$-$similarity of flow properties across different baffle lengths.
	
	The simulation results confirm the existence of the arch$-$shaped structure around the outlet for contact forces and kinetic forces, providing strong support for Beverloo's and Janda's arguments. The spatial distributions of flow properties, such as contact force, kinetic force, packing density, and grain velocity, indicate that the flow state has a global similarity exhibiting a transition between mass flow pattern and funnel flow pattern. Decreasing the baffle length causes the funnel flow pattern to expand into regions further from the outlet. Meanwhile, local self$-$similarities in packing density, grain velocity, and flow rate are again observed, consistent with Janda's findings. 
	
	The variance in baffle length has little influence on grain velocity in the local regions around the outlet; however, a similar exponential change, like the relationship between dense flow rate and baffle length, is observed for the first time. Furthermore, the dependence of the transition height from the mass flow pattern to the funnel flow pattern on the baffle length is also identified for the first time. These results demonstrate a global similarity of the flow behaviors throughout the entire hopper in terms of exponents. The earlier occurrence of the transition from the mass flow pattern to the funnel flow pattern leads to higher grain velocity, resulting in a larger flow rate. These findings indicate that a similar transition between the interior flow patterns near and far from the outlet affects the dense flow rate in the hopper. This exponential similarity also provides further insights for developing a general constitutive framework for dense granular flow.
	
	\section*{Acknowledgements}
	This work is financially supported by the National Natural Science Foundation of China (Grant No. 11574153) and the foundation of the Ministry of Industry and Information Technology of China (Grant No. TSXK2022D007).
	
	\noindent
	

\end{document}